\documentclass[conference]{IEEEtran}
\IEEEoverridecommandlockouts

\usepackage{amsmath,amssymb,amsfonts,amsthm}
\usepackage{graphicx}
\usepackage{textcomp}
\usepackage{xcolor}
\usepackage{orcidlink}
\usepackage{siunitx}
\usepackage[nolist]{acronym}
\usepackage{placeins}
\usepackage[caption=false,font=footnotesize]{subfig}
\usepackage{booktabs}
\usepackage{svg}
\usepackage[export]{adjustbox}
\usepackage{multirow}
\usepackage{float}
\usepackage{tabularx, tabulary}
\usepackage[capitalise]{cleveref}
\usepackage{algorithm}
\usepackage{algpseudocode}
\usepackage{caption}

\Crefname{equation}{Eq.}{Eqs.}
\Crefname{figure}{Fig.}{Figs.}
\Crefname{table}{Tab.}{Tabs.}

\usepackage[normalem]{ulem}
\usepackage[style=ieee,dashed=false, maxcitenames=1,mincitenames=1,eprint=false,isbn=false,url=false,doi=false,date=year]{biblatex}

\AtEveryBibitem{
    \clearfield{urlyear}
    \clearfield{urlmonth}
    \clearlist{language}
    \clearlist{issn}
}
\usepackage{tcolorbox}

\let\StandardIncludeGraphics\includegraphics%
\renewcommand{\includegraphics}[2][]{%
  \IfFileExists{#2}{%
    \StandardIncludeGraphics[#1]{#2}%
  }{%
    \IfFileExists{#2.pdf}{\StandardIncludeGraphics[#1]{#2.pdf}}{%
      \IfFileExists{#2.png}{\StandardIncludeGraphics[#1]{#2.png}}{%
        \IfFileExists{#2.jpg}{\StandardIncludeGraphics[#1]{#2.jpg}}{%
          \IfFileExists{#2.eps}{\StandardIncludeGraphics[#1]{#2.eps}}{%
            \begin{tcolorbox}[width=6cm,height=4cm,arc=0mm,auto outer arc]
            \end{tcolorbox}%
          }%
        }%
      }%
    }%
  }%
}

\def\BibTeX{{\rm B\kern-.05em{\sc i\kern-.025em b}\kern-.08em
    T\kern-.1667em\lower.7ex\hbox{E}\kern-.125emX}}

\makeatletter
\def\ps@IEEEtitlepagestyle{%
  \def\@oddfoot{\mycopyrightnotice}%
  \def\@evenfoot{}%
}
\def\mycopyrightnotice{%
	\begin{minipage}{\textwidth}
		\centering \scriptsize
		© 2026 IEEE. Personal use of this material is permitted. Permission from IEEE must be obtained for all other uses, in any current or future media, including reprinting/republishing this material for advertising or promotional purposes, creating new collective works, for resale or redistribution to servers or lists, or reuse of any copyrighted component of this work in other works.\hfill
	\end{minipage}
	\gdef\mycopyrightnotice{}
}
\makeatother

\begin{document}

\title{StreamRTPS: Increasing DDS Bandwidth Efficiency by Reducing Protocol Overhead \\
{}
}

\author{\IEEEauthorblockN{David Philipp Klüner}
\IEEEauthorblockA{\textit{Chair of Embedded Software} \\
\textit{RWTH Aachen University}\\
Aachen, Germany \\
kluener@embedded.rwth-aachen.de}
\and
\IEEEauthorblockN{Stefan Kowalewski}
\IEEEauthorblockA{\textit{Chair of Embedded Software} \\
\textit{RWTH Aachen University}\\
Aachen, Germany \\
kowalewski@embedded.rwth-aachen.de}
\and
\IEEEauthorblockN{Alexandru Kampmann}
\IEEEauthorblockA{\textit{Chair of Embedded Software} \\
\textit{RWTH Aachen University}\\
Aachen, Germany \\
kampmann@embedded.rwth-aachen.de}
}

\maketitle

\begin{acronym}\itemsep0pt
    \acro{E/E}{Electrical/Electronic}
    \acro{IVN}{In-Vehicular Network}
    \acro{CAV}{Connected and Automated Vehicle}
    \acro{CAN}{Controller Area Network}
    \acro{AV}{Automated Vehicle}
    \acro{AD}{Automated Driving}
    \acro{SDN}{Software-Defined Network}
    \acro{VM}{Virtual Machine}
    \acro{SOA}{Service-oriented Architecture}
    \acro{ROS 2}{Robot Operating System 2}
    \acro{CPS}{Cyber-Physical System}
    \acro{ECU}{Electronic Control Unit}
    \acro{AA}{Adaptive Application}
    \acro{IPC}{Inter-Process Communication}
    \acro{CM}{Communication Management}
    \acro{ROS}{Robot Operating System}
    \acro{TLS}{Transport Layer Security}
    \acro{SM}{State Management}
    \acro{EM}{Execution Management}
    \acro{OEM}{Original Equipment Manufacturer}
    \acro{ARA}{AUTOSAR Runtime for Adaptive Applications}
    \acro{OS}{Operating System}
    \acro{FC}{Function Cluster}
    \acro{FPGA}{Field Programmable Gate Array}
    \acro{GPU}{Graphics Processing Unit}
    \acro{UCM}{Update and Configuration Management}
    \acro{DL}{Deep Learning}
    \acro{AP}{Adaptive Platform}
    \acro{IDS}{Intrusion Detection System}
    \acro{S2S}{Service-to-Signal}
    \acro{MQTT}{Message Queuing Telemetry Transport}
    \acro{OTA}{Over-The-Air}
    \acro{TSN}{Time-sensitive Networking}
    \acro{UDP}{User Datagram Protocol}
    \acro{NUC}{Intel Next-Unit-of-Computing}
    \acro{TCP}{Transmission Control Protocol}
    \acro{OMG}{Object Management Group}
    \acro{QoS}{Quality of Service}
    \acro{DDS}{Data Distribution Service}
    \acro{IP}{Internet Protocol}
    \acro{SDV}{Software-Defined Vehicle}
    \acro{SPDP}{Simple Participant Discovery Protocol}
    \acro{SEDP}{Simple Endpoint Discovery Protocol}
    \acro{DMA}{Direct Memory Access}
    \acro{HMI}{Human Machine Interface}
    \acro{RTPS}{Real-Time Publish Subscribe}
    \acro{ML}{Machine Learning}
    \acro{API}{Application Programming Interface}
    \acro{IMU}{Inertial Measurement Unit}
    \acro{HPC}{High-Performance Computer}
    \acro{Wi-Fi}{Wireless Fidelity}
    \acro{V2X}{Vehicle-to-Everything}
    \acro{4G}{4G}
    \acro{NIC}{Network Interface Card}
    \acro{NACK}{Negative Acknowledgement}
\end{acronym}

\begin{abstract}

In this paper, we propose three extensions to the \acl{RTPS} wire protocol, on which \ac{DDS} is based, to improve bandwidth efficiency. 
First, a stream negotiation mechanism exchanges static header information during discovery, replacing the full RTPS header at runtime with a compact \SI{2}{\byte} identifier. 
Second, a payload aggregation scheme aggregates samples for the same locator into single UDP packets, reducing IP and UDP header costs.
Third, a predictive heartbeat suppression strategy reduces control traffic by omitting heartbeats for periodic communication patterns, falling back upon detected loss or timing violations. 
All three mechanisms preserve \ac{RTPS} compatibility by extending DDS discovery to activate these features when supported. 
Experimental results show that stream headers reduce bandwidth consumption by up to \SI{27.9}{\percent} compared to conventional RTPS under best-effort transport, and that heartbeat suppression yields a further \SI{22.7}{\percent} reduction on top of stream headers under reliable transport, while preserving transmission latency in both cases.

Implementation available at \url{https://github.com/embedded-software-laboratory/StreamRTPS}.
\end{abstract}


\section{Introduction}
\label{sec:intro}

Communication protocols, such as TCP or QUIC, commonly prepend a header to every transmitted unit of payload \cite{floyd_tcp_1994, langley_quic_2017}.
Such headers typically carry metadata including sequence numbers, address information, and fields to verify transmission correctness.
Although essential for reliable and structured communication, these headers impose a constant per-packet overhead, decreasing bandwidth efficiency \cite{bormann_robust_2001}.

This overhead is pronounced in \ac{RTPS}, the protocol underlying the widely used \ac{DDS} standard.
The header of \ac{RTPS} is largely static and contains substantial redundant information in successive packets of the same data stream.
For small payloads, in particular, the cumulative header cost can dominate the total transmission size and lead to low bandwidth efficiencies.

\begin{figure}[tbp!]
    \centering
    Mean Bandwidth Comparison\par\medskip
    \includegraphics[width=0.48\textwidth]{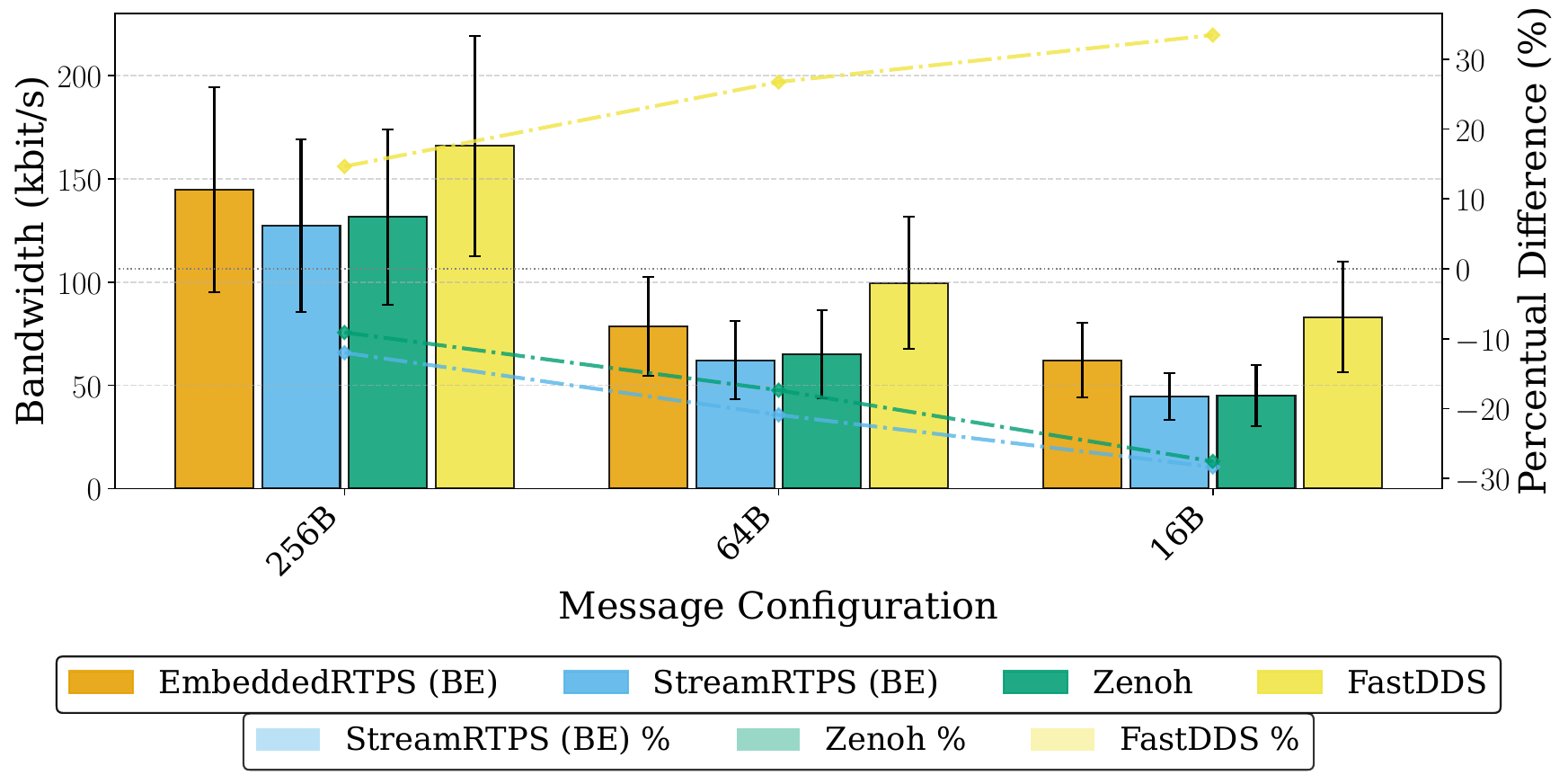}
    \caption{Results comparing FastDDS, StreamRTPS, EmbeddedRTPS and Zenoh in a 4:2-4:2 topology with two senders and two receivers, two topics per sender at \SI{25}{\hertz}. Error bars denote one standard deviation.
    }
    \label{fig:motivation_result}
\end{figure}


Modern middlewares, such as Zenoh, possess more compact wire formats \cite{chisalita_stepping_2025, baron_performance_2025}.
We compared two \ac{RTPS} implementations (FastDDS and EmbeddedRTPS) to Zenoh with identical payloads and measured the total bandwidth of each.
As shown in \cref{fig:motivation_result}, both \ac{RTPS} implementations incur higher overhead than Zenoh across all payload sizes.
Specifically in bandwidth-constrained networks, such as industrial, automotive, or robotics, in which DDS is often employed, bandwidth efficiency is an important factor.
With increasing bandwidth demands of applications, wasting bandwidth on unnecessary header transmissions is not desirable in IoT, automotive or robotics applications \cite{zhu_requirements-driven_2021, baron_performance_2025}.

To address this inefficiency, we propose a backward-compatible extension to the \ac{RTPS} wire protocol, termed \emph{StreamRTPS} (SRTPS).
StreamRTPS is guided by three objectives:
First, moving static header information to a single handshake ahead of communication instead of repeating them in every packet. 
Second, aggregating packets meant for the same locator to reduce transport layer overhead of multiple small transmissions.
Third, reducing control traffic in reliable transport, where possible, without loss of reliability.

Consequently, the main contributions of this paper are:
\begin{enumerate} 
    \item Presentation of our StreamRTPS method to reduce RTPS header overhead, message aggregation techniques, and control-traffic reduction.
    \item An open-source implementation integrated into EmbeddedRTPS.
    \item Evaluation and Interpretation of performance and latency characteristics for varying packet sizes, and data rates.
\end{enumerate}

The paper is structured as follows:
\Cref{sec:related} presents other research on DDS middleware related to the protocol.
\Cref{sec:background} provides an introduction to middleware and distributed systems. 
\Cref{sec:method} presents our approach to reduce bandwidth-overhead in RTPS transmissions.
\Cref{sec:evaluation} describes our experimental setup and parametrization.
In \Cref{sec:results}, we present the results of our experiments and interpret them.
Finally, \Cref{sec:conclusion} presents our conclusion and the applicability of our method.

\section{Related Work}
\label{sec:related}
Header overhead reduction is a known problem in networking. Protocols such as ROHC~\cite{bormann_robust_2001} negotiate shared context during setup and transmit compressed headers at runtime. However, these approaches rely on per-packet compression and decompression, introducing variable processing latency in contrast to our approach.
We therefore review related work along two axes: alternative middlewares with compact wire formats, and DDS-specific optimization efforts.

\subsection{Middlewares}
Our work addresses issues relative to two middlewares whose compact wire formats motivate the overhead reduction targeted in this paper.
Zenoh is a publish/subscribe/query protocol by \citeauthor{corsaro_zenoh_2023} \cite{corsaro_zenoh_2023} that unifies data in motion, data at rest, and computations. All operations reduce to three primitives (\texttt{put}, \texttt{delete}, \texttt{get}) over key/value resources. Zenoh can operate on OSI Layer~2 with approximately \SI{5}{\byte} overhead per message, targeting devices down to 8-bit microcontrollers.
MQTT is a broker-centric publish/subscribe protocol standardized by OASIS for IoT \cite{sengul_message_2023}. Clients communicate through a central broker over TCP/IP with a \SI{2}{\byte} fixed header. Three \ac{QoS} levels provide at-most-once, at-least-once, and exactly-once delivery.

\subsection{DDS}
Despite widespread deployment of \ac{DDS} in robotics, automotive, and defense, research directly addressing the \ac{RTPS} wire protocol remains sparse.
Several studies target \ac{SPDP}/\ac{SEDP} efficiency. 
Bloom-filter-based discovery was proposed by \citeauthor{sanchez-monedero_bloom_2011}~\cite{sanchez-monedero_bloom_2011} and extended with dynamic filters by \citeauthor{khaefi_node_2014}~\cite{khaefi_node_2014}.
These reduce discovery traffic but leave the data transmission path unchanged.

Application-level optimizations have been proposed primarily for \acl{V2X}. 
\citeauthor{peeck_protocol_2023}~\cite{peeck_protocol_2023} propose an efficient fragment retransmission scheme for lossy channels. \citeauthor{sperling_reducing_2024}~\cite{sperling_reducing_2024} took a subscriber-centric approach, distributing only relevant parts of an image to each receiver.
These works operate on top of an unmodified \ac{RTPS} wire protocol.
\citeauthor{bode_adopting_2024}~\cite{bode_adopting_2024} introduced user-space networking for DDS via DPDK and XDP extensions, based on a systematic implementation evaluation~\cite{bode_dds_2023}. While this accelerates packet processing, it does not reduce the data transmitted on the wire.

To our knowledge, no prior work has modified the \ac{RTPS} wire format itself. Our approach provides an alternative data path that supersedes the standard message format for the majority of transmission cases. In contrast to existing header reduction techniques, integrating directly with \ac{RTPS} promises maximum efficiency while preserving \ac{RTPS}'s real-time properties, important for its application domains.

\begin{figure*}[tpb!]
    \centering
    \includegraphics[width=1.0\textwidth]{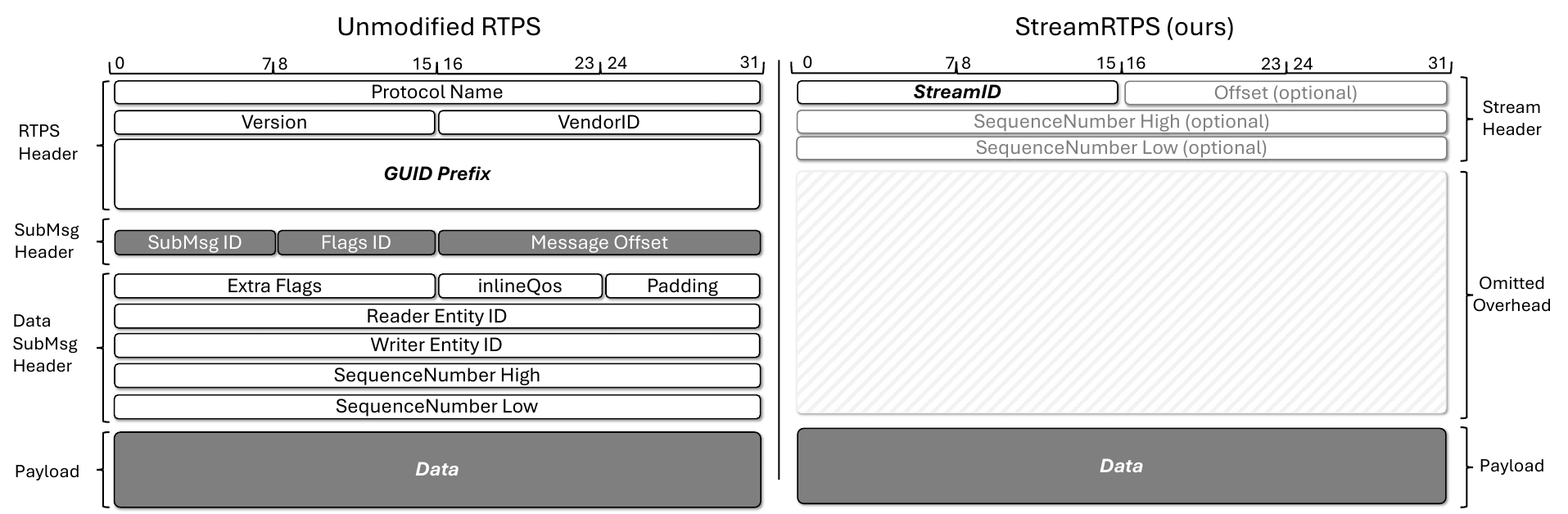}
    \caption{Illustration of an unmodified RTPS packet and our StreamRTPS packet. In the Simple format only the StreamID is required for transmission, while the Standard and Reliable formats add the offset field for packet aggregation and sequence numbers for reliable transmissions.}
    \label{fig:rtps_header}
\end{figure*}

\section{Background}
\label{sec:background}

\ac{DDS} is an \ac{OMG} middleware standard for distributed real-time publish-subscribe communication. Data is organized into typed topics to which publishers write samples and subscribers register interest. DDS is the default middleware in \ac{ROS 2} and widely adopted in robotics and automotive systems. At the transport level, it relies on the \ac{RTPS} wire protocol over \ac{UDP}, \ac{TCP}, or shared memory, and provides configurable \ac{QoS} policies for reliability, history, and resource limits.
\ac{RTPS} is the interoperability wire protocol of DDS, governing discovery and data transmission.
Discovery proceeds in two multicast stages. The \ac{SPDP} advertises participant presence through periodic announcements. Subsequently, the \ac{SEDP} exchanges endpoint metadata (topic names, types, \ac{QoS}) between discovered participants, establishing matched communication pairs.
User data is carried in RTPS messages consisting of a fixed header followed by one or more submessages. \texttt{Data} submessages carry samples with sequence numbers for ordering, and a single message may bundle submessages for different readers. Reliable delivery is achieved through \texttt{Heartbeat} submessages, in which writers advertise available sequence number ranges, and \texttt{AckNack} submessages, in which readers request retransmission of missing samples.

\section{Methodology}
\label{sec:method}
We propose three extensions to the Real-Time Publish-Subscribe (RTPS) wire protocol aimed at improving bandwidth efficiency:
First, a stream negotiation mechanism that reduces per-packet header overhead by establishing shared context between endpoints.
Second, a deadline-aware message aggregation scheme that amortizes IP and UDP header costs across multiple samples.
Third, a predictive heartbeat suppression strategy that reduces control traffic for periodic communication patterns.

\subsection{Stream-based Transport}
\label{sec:method:streams}
Currently, the unmodified RTPS message and submessage headers introduce a combined fixed overhead of \SI{44}{\byte} per packet, as shown in \cref{fig:rtps_header}.
However, the majority of these header fields, such as the protocol identifier, version, vendor ID and GUID prefix, remain static throughout the lifetime of a communication session. 
We exploit this observation by exchanging this static header information during discovery and replacing the full RTPS header with a compact stream header at runtime. 
The stream header encodes only strictly necessary runtime-dynamic information and is presented with its negotiation procedure in the next paragraph.

\begin{figure}[bp!]
    \centering
    \includegraphics[width=0.5\textwidth]{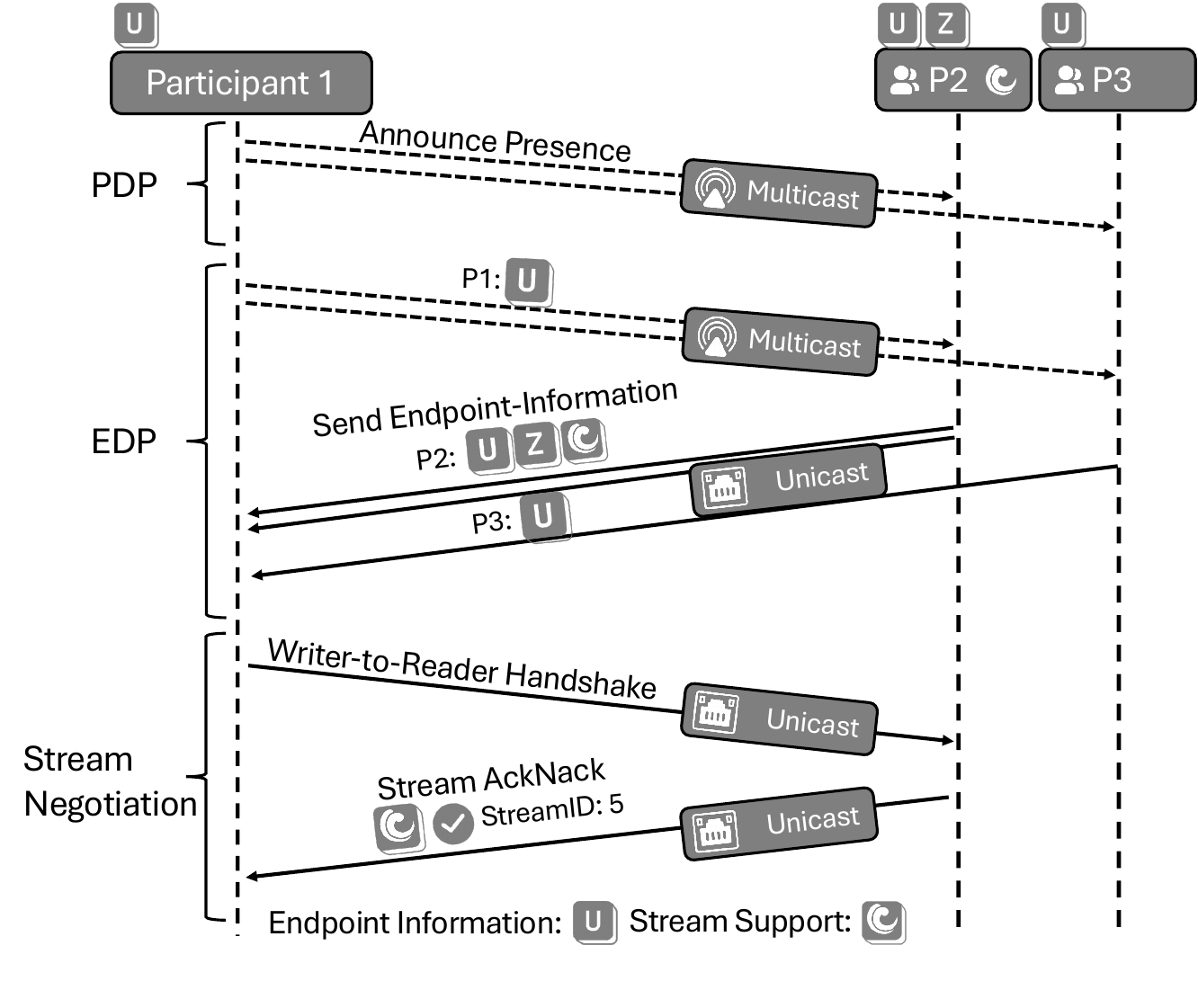}
    \caption{Illustration of our SPDP, SEDP and our new stream negotiation handshake. In this handshake, transmission parameters and the resulting StreamID are exchanged.}
    \label{fig:stream_negotiation}
\end{figure}

\subsubsection{Stream Header}
\label{sec:method:stream_header}
The proposed stream header exists in three formats of increasing capability, each a prefix of the next:
\begin{itemize}
    \item \textbf{Simple} (\SI{2}{\byte}): Contains only the \texttt{StreamID} (\texttt{uint16}), a reader-unique identifier referencing the full header context negotiated during discovery. Used for best-effort streams without aggregation.
    \item \textbf{Standard} (\SI{4}{\byte}): Adds an \texttt{Offset} field (\texttt{uint16}), a byte offset to the next stream packet within the same UDP datagram. Enables aggregating messages from different writers destined for the same locator, serving a role similar to the existing submessage mechanism with less overhead.
    \item \textbf{Reliable} (\SI{12}{\byte}): Adds a \texttt{SequenceNumber} (\SI{8}{\byte}), required for reliable communication to preserve sample ordering and enable retransmission.
\end{itemize}
Receivers distinguish stream packets from standard RTPS messages by checking whether the first four bytes match the \texttt{RTPS} protocol identifier, stream packets do not carry this prefix.

\subsubsection{Stream Negotiation}
\label{sec:method:negotiation}

To establish stream sessions, we introduce an additional pair of built-in endpoints and a per-participant handler: the Stream Negotiation (STNE) writer, reader, and handler.
The negotiation procedure, shown in \cref{fig:stream_negotiation}, is initiated conditionally after the standard Simple Endpoint Discovery Protocol (SEDP) phase has completed.
We extended SEDP-exchanged writer and reader information to include flags indicating support for streams and message aggregation. If these flags are present and set, the stream negotiation proceeds as follows.
The writer selects a stream header format based on the mutually supported capabilities: if both endpoints support only streams, the Simple format is used. 
If both additionally support aggregation, the Standard format is selected. Reliable writers always use the Reliable format.
The writer transmits its static header information together with the chosen format to the corresponding STNE handler on the reader side.
The reader's STNE handler validates the request, assigns a \texttt{StreamID}, and responds with an acknowledgment containing this identifier.
\texttt{StreamIDs} are unique per reader of a receiving participant to ensure that the receiving socket can distinguish which topic the packet belongs to.
The current version does not support multicast delivery, as a single packet cannot carry distinct \texttt{StreamIDs} for multiple readers. Multicast traffic therefore falls back to standard RTPS. Future work will extend the mechanism with shared \texttt{StreamIDs} in a separate address range for multicast communication.
Subsequent data transmissions then use the negotiated \texttt{StreamID} in place of the full RTPS header, reducing wire overhead per message from \SI{44}{\byte} to \SI{2}{\byte} (Simple), \SI{4}{\byte} (Standard), or \SI{12}{\byte} (Reliable).
If the STNE handler denies the request or does not respond within 3 retries over \SI{20}{\milli\second}, the writer falls back to regular RTPS communication, retaining compatibility with the RTPS standard.

\subsection{Cross-Writer Payload Aggregation}
\label{sec:method:aggregation}
In standard \ac{RTPS}, each payload is typically transmitted as a separate UDP datagram, incurring the full cost of IP and UDP headers per sample.
On bandwidth-constrained networks, this per-packet overhead can represent a significant fraction of the total traffic, particularly for small, high-frequency signals common in fast control loops.
We mitigate IP and UDP header overhead by aggregating multiple samples destined for the same transport locator into a single UDP datagram, distributing the header cost across all bundled samples.
Within the proposed stream header format (Section~\ref{sec:method:stream_header}), the \texttt{Offset} field allows for the chaining of several packets within an aggregated datagram so that the receiver can distinguish them.

\begin{figure}[bp]
    \centering
    \includegraphics[width=0.5\textwidth]{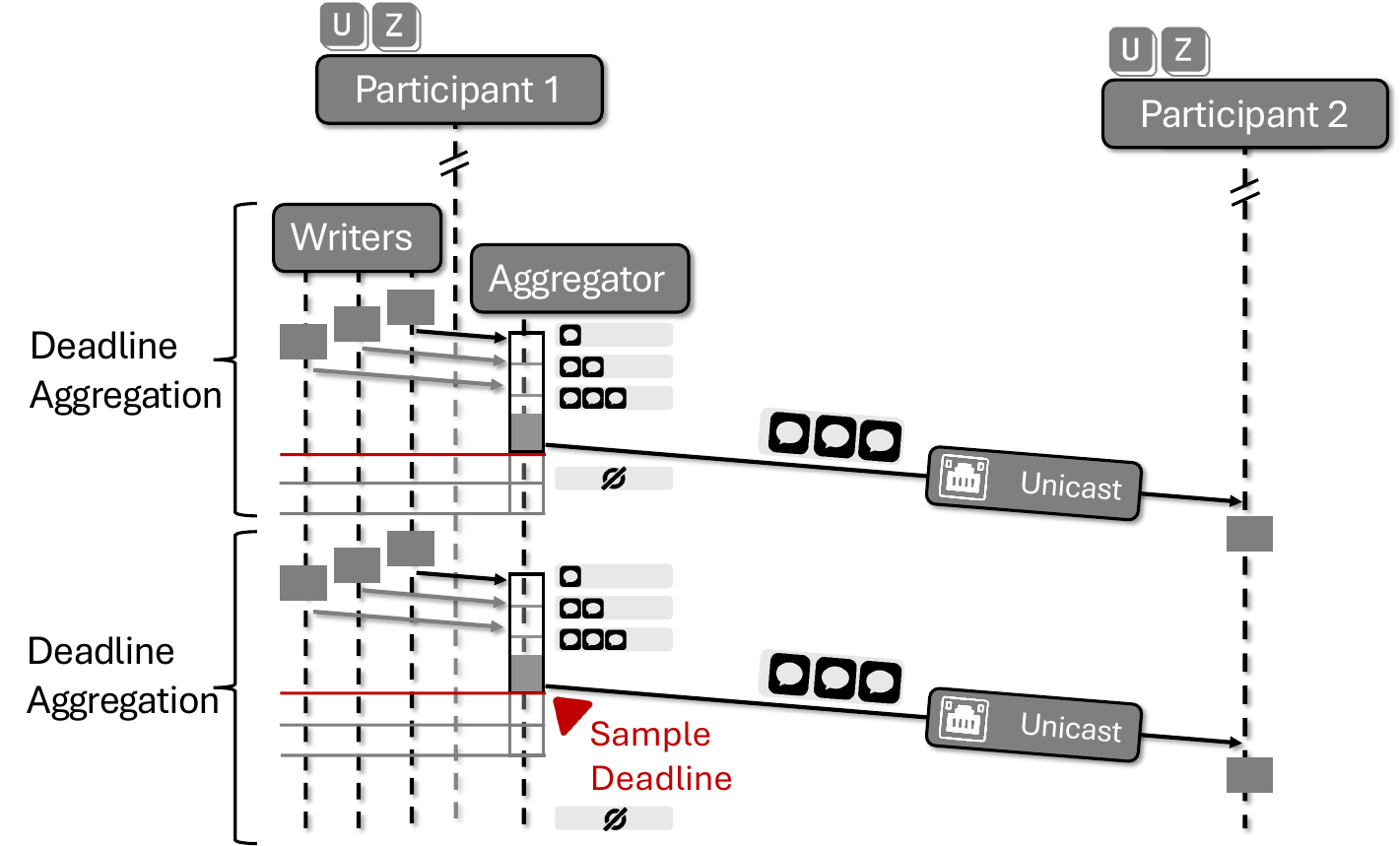}
    \caption{Illustration of Deadline-based Payload Aggregation. Samples are aggregated to the earliest deadline of a given sample, then sent and the buffer emptied.}
    \label{fig:aggregation}
\end{figure}

Since writers on distinct topics may produce samples at different rates, a fixed time window risks either delaying latency-sensitive samples or flushing too early.
We therefore tie the aggregation window to the QoS deadlines of pending samples, as illustrated in \cref{fig:aggregation}.
When a sample arrives, it is appended to a per-locator buffer.
A dedicated send loop flushes the buffer at the earliest deadline among all pending samples, bundling everything that arrived in the interim into a single datagram.
If a new sample arrives with a deadline earlier than the currently scheduled flush time, the buffer is flushed immediately.
This ensures that deadlines are never violated while maximizing aggregation, at the cost of increasing transmission latency up to the configured deadline.

\subsection{Heartbeat Suppression}
\label{sec:method:traffic_implied}
Standard RTPS reliability requires writers to send periodic heartbeats so that readers can detect missing samples.
For periodic writers, these heartbeats are redundant, as receivers can infer loss from a gap in the expected arrival pattern.
Instead of waiting for a heartbeat to verify that a message was lost, the reader can detect loss directly when receiving the next message.
We exploit this with a new heartbeat mode for periodic communication in which the reader monitors the arrival pattern and the writer can in turn suspend or significantly delay heartbeat transmission.
The algorithm for this is shown in \cref{alg:hb-suppression}, with parameters listed in \cref{tab:hb-params}.

\begin{algorithm}[t]
\caption{Heartbeat Suppression}\label{alg:hb-suppression}
\begin{algorithmic}[1]
\Statex \textbf{State:} mode $m \gets \textsc{Acquire}$,
    interval ring buffer $I$,
    period $T \gets T_0$, NACK streak $c \gets 0$
\Statex
\Procedure{OnNewChange}{}
    \State push $(t_{\mathrm{now}} - t_{\mathrm{prev}})$ into $I$
    \If{$|I| \ge N_{\min}$ \textbf{and}
        $\operatorname{Var}(I) < \sigma^2_{\max}$}
        \State $m \gets \textsc{Implicit}$;\;
            $T \gets T_{\mathrm{impl}}$
            \hfill $//$ \textit{suppress}
    \Else
        \State $m \gets \textsc{Explicit}$;\;
            $T \gets \mu(I) \cdot M$
    \EndIf
\EndProcedure
\Statex
\Procedure{OnAckNack}{hadNacks}
    \If{\textbf{not} hadNacks}
        \State $c \gets 0$
    \ElsIf{$c{+}1 \ge K$}
        \State $m \gets \textsc{Acquire}$;\; reset $I$;\; $T \gets T_0$
    \Else
        \State $c \mathrel{+}= 1$;\;
            $T \gets T/2$;\; $m \gets \textsc{Explicit}$
    \EndIf
\EndProcedure
\Statex
\Procedure{ShouldSend}{}
    \State \Return $t_{\mathrm{now}} - t_{\mathrm{lastHB}} \ge T$
\EndProcedure
\end{algorithmic}
\end{algorithm}

\paragraph{Writer-side heartbeat suppression}
To initially detect the current send pattern, each writer maintains a ring buffer $I$ of recent inter-sample intervals and a mode $m \in \{\textsc{Acquire}, \textsc{Explicit}, \textsc{Implicit}\}$.
Initially, the writer begins operation in \textsc{Acquire} mode.
On every new sample, \textsc{OnNewChange} pushes the latest interval into $I$ and, once at least $N_{\min}$ intervals are collected, evaluates periodicity.
When the variance $\operatorname{Var}(I)$ is below the $ \sigma^2_{\max}$ threshold, the writer enters \textsc{Implicit} mode, extending the heartbeat period to $T_{\mathrm{impl}}$.
If the variance exceeds the threshold, \textsc{Explicit} mode sets the period to $\mu(I) \cdot M$, where $\mu(I)$ is the sample mean and $M$ a target messages-per-heartbeat count, spacing heartbeats proportionally to the observed rate.
In \textsc{Implicit}, a fallback heartbeat is emitted in $T_{\mathrm{impl}}$ as a backup. 
On \acp{NACK}, \textsc{OnAckNack} halves the period to respond to detected loss and after $K$ consecutive \acp{NACK} resets to \textsc{Acquire}.

\paragraph{Reader-side proactive loss detection}
Suppression maintains safety because the reader independently rate-locks onto the sample stream using the same periodicity criterion.
Once locked, if the gap since the last sample exceeds $\alpha \cdot \mu(I)$, the reader fires a proactive \ac{NACK} immediately.
For a \SI{10}{\milli\second} sample period, this detects loss within approximately \SI{30}{\milli\second}, well before the suppressed heartbeat at $T_{\mathrm{impl}} = \SI{2000}{\milli\second}$, resulting in faster loss detection than the standard implementation.
Note that proactive detection is faster than the standard heartbeat period $T_0$ only when $\alpha \cdot \mu(I) < T_0$ for sample periods below $T_0 / \alpha \approx \SI{83}{\milli\second}$ with our parameters.

\section{Evaluation}
\label{sec:evaluation}
We evaluated the three proposed extensions in separate experiments:
\begin{enumerate}
    \item \textbf{StreamRTPS}: Compares standard RTPS with StreamRTPS to quantify stream header overhead reduction.
    \item \textbf{Payload Aggregation}: Measures the effect of deadline-aware aggregation on bandwidth and latency.
    \item \textbf{Heartbeat Suppression}: Evaluates heartbeat suppression under reliable transport, including loss recovery.
\end{enumerate}

\begin{table}[t]
\centering
\caption{Heartbeat suppression parameters for \cref{alg:hb-suppression}.}
\label{tab:hb-params}
\begin{tabular}{llr}
    \toprule
    \textbf{Parameter} & \textbf{Symbol} & \textbf{Value} \\
    \midrule
    Window size               & $N_{\min}$            & 5 \\
    Max.\ variance            & $\sigma^2_{\max}$     & \SI{400}{\micro\second\squared} \\
    Implicit heartbeat period & $T_{\mathrm{impl}}$   & \SI{2000}{\milli\second} \\
    Standard heartbeat period & $T_0$                 & \SI{250}{\milli\second} \\
    NACK streak threshold     & $K$                   & 3 \\
    Messages per heartbeat    & $M$                   & 5 \\
    Proactive NACK multiplier & $\alpha$              & 3 \\
    \bottomrule
\end{tabular}
\end{table}

\subsection{Hardware and Software Setup}
\label{sec:eval:setup}
Our testbed consists of four Lenovo ThinkCentre M900 small-form factor machines, each with an Intel Core i5-6500T processor, \SI{16}{\giga\byte} of DDR4 RAM and an M.2 SSD, connected via a central switch over \SI{1}{\giga\bit\per\second} full-duplex Ethernet.
The machines run Ubuntu 24.04.3 LTS with the GNU/Linux 6.17.0-14-generic x86\_64 kernel. No additional real-time configuration was applied to the Linux installation.
As baselines we use EmbeddedRTPS~\cite{kampmann_portable_2019} and FastDDS~v3.4. We also included Zenoh~v1.1.0 for our motivating example.
Our proposed StreamRTPS extensions are implemented as a fork of EmbeddedRTPS.
Bandwidth was captured with TShark, covering both unicast traffic between testbed machines and multicast traffic. All bandwidth values are reported as the per-sender egress rate captured at the sender's NIC. We used LTTng for kernel and userspace tracing, and PTP for clock synchronization with a measured accuracy better than \SI{25}{\micro\second}.

\subsection{Experiment Parametrization}
\label{sec:eval:params}
We conducted all experiments by measuring 500 steady-state samples at 25 Hz send rate across five separate executions.
Experiments for Stream transport and message aggregation are conducted with best-effort (BE) reliability, while experiments modifying heartbeats are conducted with reliable transport (RL). 
Unless specified, experiments use 2 sender machines and 2 receiver machines in total with two subscribers or two publishers per machine, denoted as 4:2-4:2 with each subscriber-publisher pair using a single topic. 
We vary the message sizes between seven payload sizes: \SI{16}{\byte}, \SI{32}{\byte}, \SI{64}{\byte}, \SI{128}{\byte}, \SI{256}{\byte}, \SI{512}{\byte}, and \SI{1024}{\byte}.
For aggregation experiments, the timing of samples and their deadlines is important. 
Consequently, we divide each transmission period into four equal time slots. For aggregation experiments, the notation $n_1$-$n_2$-$n_3$-$n_4$ indicates the number of publishers transmitting in each slot. For example, 1-1-1-0 at \SI{25}{\hertz} (\SI{40}{\milli\second} period) indicates three publish events separated by \SI{10}{\milli\second}. When payload aggregation is enabled, the aggregation deadline is specified in each figure.

\begin{figure*}[!btp]
\centering

StreamRTPS Bandwidth Measurements\par\medskip
\makebox[\textwidth]{%
\begin{minipage}[t]{0.48\textwidth}
\vspace{0pt}
\centering
\includegraphics[width=\textwidth]{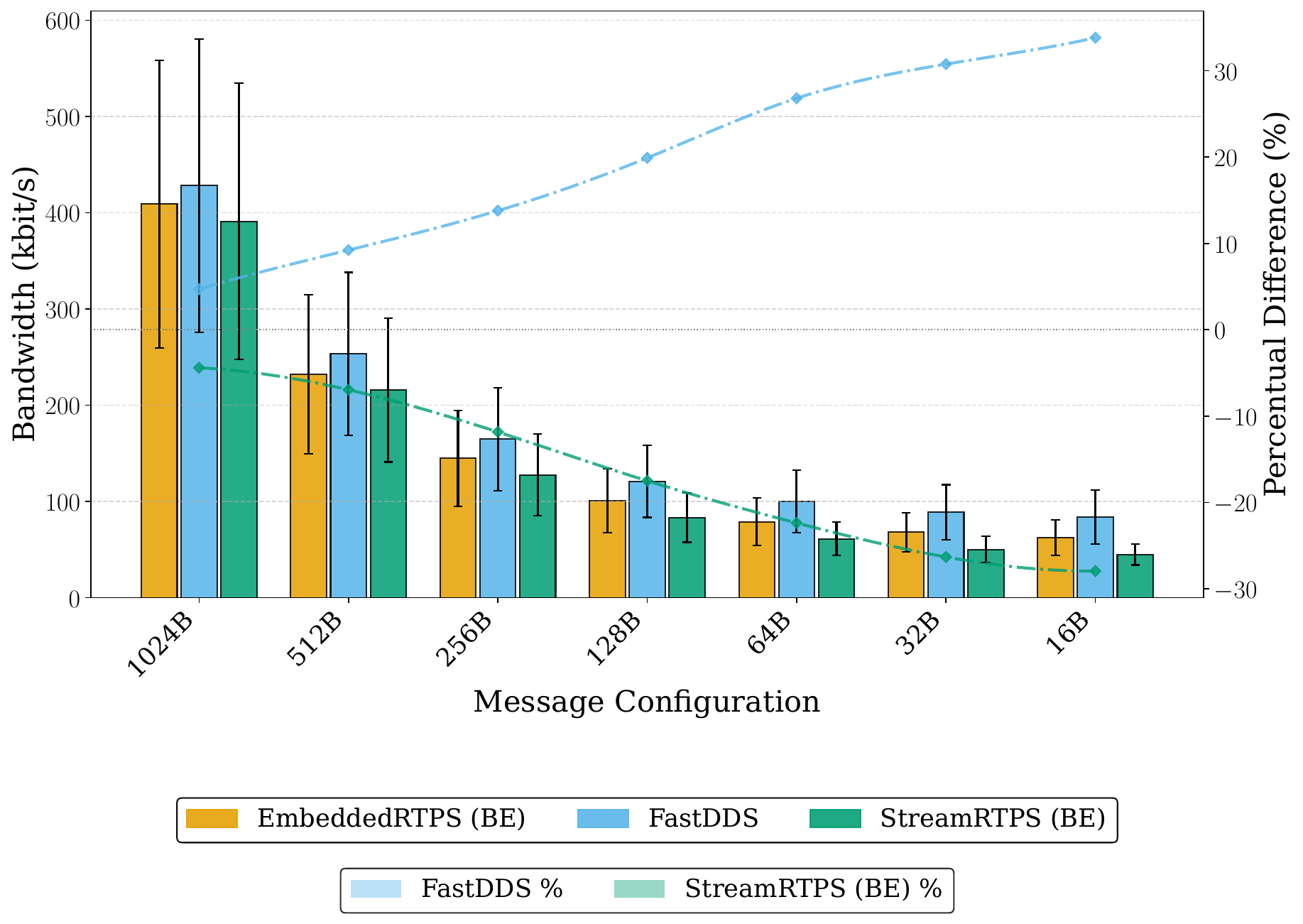}

\end{minipage}%
\hfill%
\begin{minipage}[t]{0.48\textwidth}
\vspace{0pt}
\centering
\resizebox{\textwidth}{!}{
\begin{tabular}{lccc}
    \toprule
    \textbf{Payload} & \textbf{FastDDS} & \textbf{EmbeddedRTPS} & \textbf{SRTPS} \\
    \midrule
    \multicolumn{4}{l}{\textit{Mean Bandwidth $B$ [\SI{}{\kilo\bit\per\second}]}} \\
    \hspace{1mm} \SI{16}{\byte}   & 83.5 $\pm$ 28.1 (\SI{+33.8}{\percent}) & 62.4 $\pm$ 18.6  & 45.0 $\pm$ 11.0 (\textbf{\SI{-27.9}{\percent}}) \\
    \hspace{1mm} \SI{32}{\byte}   & 88.9 $\pm$ 28.5 (\SI{+30.8}{\percent}) & 68.0 $\pm$ 20.3  & 50.1 $\pm$ 13.6 (\textbf{\SI{-26.3}{\percent}}) \\
    \hspace{1mm} \SI{64}{\byte}   & 100.0 $\pm$ 32.7 (\SI{+26.8}{\percent}) & 78.9 $\pm$ 24.6  & 61.2 $\pm$ 17.5 (\textbf{\SI{-22.4}{\percent}}) \\
    \hspace{1mm} \SI{128}{\byte}  & 120.9 $\pm$ 37.5 (\SI{+19.9}{\percent}) & 100.9 $\pm$ 33.2  & 83.2 $\pm$ 25.6 (\textbf{\SI{-17.5}{\percent}}) \\
    \hspace{1mm} \SI{256}{\byte}  & 164.7 $\pm$ 53.3 (\SI{+13.8}{\percent}) & 144.8 $\pm$ 49.7  & 127.6 $\pm$ 42.4 (\textbf{\SI{-11.8}{\percent}}) \\
    \hspace{1mm} \SI{512}{\byte}  & 253.4 $\pm$ 84.6 (\SI{+9.2}{\percent}) & 232.0 $\pm$ 82.8  & 215.9 $\pm$ 74.9 (\textbf{\SI{-7.0}{\percent}}) \\
    \hspace{1mm} \SI{1024}{\byte} & 428.2 $\pm$ 152.3 (\SI{+4.7}{\percent}) & 409.0 $\pm$ 149.5  & 391.0 $\pm$ 143.5 (\textbf{\SI{-4.4}{\percent}}) \\
    \midrule
    \multicolumn{4}{l}{\textit{Mean Transmission Latency $L_T$ [\SI{}{\milli\second}]}} \\
    \hspace{1mm} \SI{16}{\byte}   & 0.53 $\pm$ 0.09 (\SI{-21.1}{\percent}) & 0.67 $\pm$ 0.12  & 0.69 $\pm$ 0.19 (\textbf{\SI{+3.0}{\percent}}) \\
    \hspace{1mm} \SI{32}{\byte}   & 0.54 $\pm$ 0.10 (\SI{-18.9}{\percent}) & 0.66 $\pm$ 0.12  & 0.66 $\pm$ 0.17 (\textbf{\SI{-0.1}{\percent}}) \\
    \hspace{1mm} \SI{64}{\byte}   & 0.52 $\pm$ 0.10 (\SI{-24.0}{\percent}) & 0.68 $\pm$ 0.10  & 0.67 $\pm$ 0.10 (\textbf{\SI{-1.3}{\percent}}) \\
    \hspace{1mm} \SI{128}{\byte}  & 0.52 $\pm$ 0.10 (\SI{-22.3}{\percent}) & 0.66 $\pm$ 0.11  & 0.68 $\pm$ 0.10 (\textbf{\SI{+3.1}{\percent}}) \\
    \hspace{1mm} \SI{256}{\byte}  & 0.52 $\pm$ 0.10 (\SI{-24.2}{\percent}) & 0.68 $\pm$ 0.12  & 0.69 $\pm$ 0.11 (\textbf{\SI{+0.4}{\percent}}) \\
    \hspace{1mm} \SI{512}{\byte}  & 0.52 $\pm$ 0.11 (\SI{-25.0}{\percent}) & 0.69 $\pm$ 0.16  & 0.66 $\pm$ 0.17 (\textbf{\SI{-5.2}{\percent}}) \\
    \hspace{1mm} \SI{1024}{\byte} & 0.55 $\pm$ 0.10 (\SI{-21.6}{\percent}) & 0.70 $\pm$ 0.14  & 0.68 $\pm$ 0.09 (\textbf{\SI{-3.0}{\percent}}) \\
    \bottomrule
\end{tabular}
}
\end{minipage}%
}
\caption{Table and illustration showing mean bandwidth and transmission latency for StreamRTPS, FastDDS and EmbeddedRTPS. Error bars in the figure and error margins in the table denote one standard deviation.}
\label{fig:results:streams}
\end{figure*}
\begin{figure*}[!tp]
\centering
\makebox[\textwidth]{%
\begin{minipage}[t]{0.48\textwidth}
\caption*{StreamRTPS Payload Aggregation Measurements}
\vspace{0pt}
\centering
\resizebox{\textwidth}{!}{ 
\begin{tabular}{lccc}
    \toprule
    \textbf{Deadline} & \textbf{EmbeddedRTPS} & \textbf{SRTPS} & \textbf{SRTPS Aggregation} \\
    \midrule
    \multicolumn{4}{l}{\textit{Mean Bandwidth $B$ [\SI{}{\kilo\bit\per\second}]}} \\
    \hspace{1mm} \SI{2}{\milli\second}  & 323.2 $\pm$ 119.2  & 252.2 $\pm$ 92.2 (\SI{-22}{\percent}) & 226.4 $\pm$ 82.7 (\textbf{\SI{-30}{\percent}}) \\
    \hspace{1mm} \SI{8}{\milli\second}  & 323.4 $\pm$ 120.0  & 252.7 $\pm$ 91.9 (\SI{-22}{\percent}) & 226.1 $\pm$ 81.9 (\textbf{\SI{-30}{\percent}}) \\
    \hspace{1mm} \SI{12}{\milli\second} & 323.7 $\pm$ 119.1  & 252.2 $\pm$ 91.6 (\SI{-22}{\percent}) & 211.9 $\pm$ 75.6 (\textbf{\SI{-35}{\percent}}) \\
    \hspace{1mm} \SI{18}{\milli\second} & 323.5 $\pm$ 119.4  & 252.5 $\pm$ 91.8 (\SI{-22}{\percent}) & 211.4 $\pm$ 76.9 (\textbf{\SI{-35}{\percent}}) \\
    \hspace{1mm} \SI{22}{\milli\second} & 323.3 $\pm$ 119.0  & 252.6 $\pm$ 92.7 (\SI{-22}{\percent}) & 206.8 $\pm$ 75.3 (\textbf{\SI{-36}{\percent}}) \\
    \midrule
    \multicolumn{4}{l}{\textit{Mean Transmission Latency $L_T$ [\SI{}{\milli\second}]}} \\
    \hspace{1mm} \SI{2}{\milli\second}  & 0.70 $\pm$ 0.09  & 0.66 $\pm$ 0.10 (\SI{-3}{\percent}) & 2.96 $\pm$ 0.38 (\textbf{\SI{+330}{\percent}}) \\
    \hspace{1mm} \SI{8}{\milli\second}  & 0.69 $\pm$ 0.10  & 0.67 $\pm$ 0.14 (\SI{-3}{\percent}) & 8.94 $\pm$ 0.42 (\textbf{\SI{+1201}{\percent}}) \\
    \hspace{1mm} \SI{12}{\milli\second} & 0.66 $\pm$ 0.09  & 0.72 $\pm$ 0.16 (\SI{+5}{\percent}) & 8.05 $\pm$ 4.92 (\textbf{\SI{+1070}{\percent}}) \\
    \hspace{1mm} \SI{18}{\milli\second} & 0.67 $\pm$ 0.09  & 0.71 $\pm$ 0.17 (\SI{+3}{\percent}) & 13.67 $\pm$ 4.93 (\textbf{\SI{+1888}{\percent}}) \\
    \hspace{1mm} \SI{22}{\milli\second} & 0.68 $\pm$ 0.10  & 0.67 $\pm$ 0.10 (\SI{-3}{\percent}) & 13.18 $\pm$ 7.96 (\textbf{\SI{+1818}{\percent}}) \\
    \bottomrule
\end{tabular}
}
\end{minipage}%
\hfill%
\begin{minipage}[t]{0.48\textwidth}
\caption*{Payload Aggregation Measurements for varying Deadlines}
\vspace{0pt}
\centering
\includegraphics[width=\textwidth]{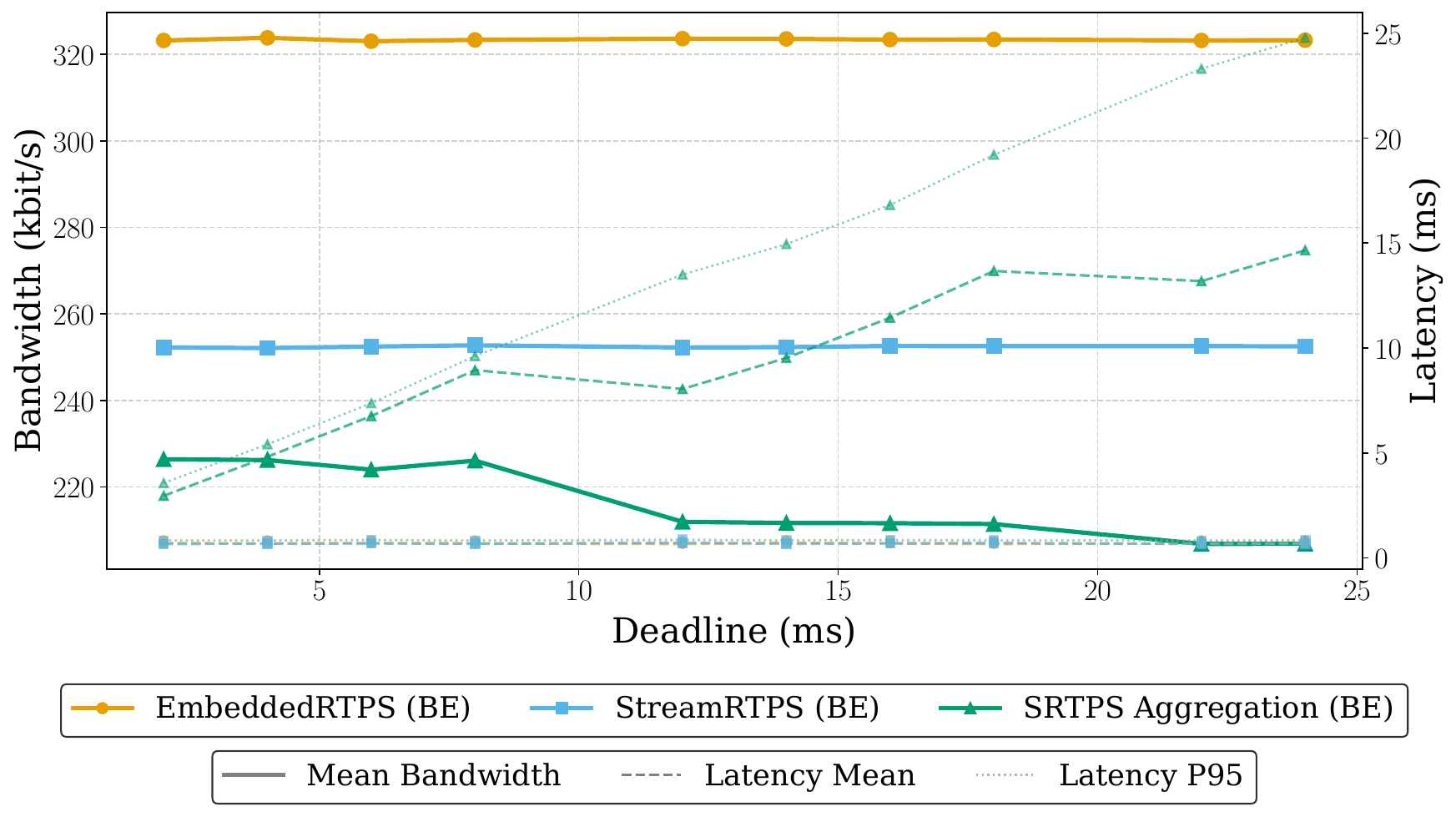}

\end{minipage}%
}
\caption{Table and illustration showing the absolute bandwidth and transmission latency for our payload aggregation approach with 128 byte transmissions in a 8:1-8:1 topology and \SI{10}{\milli\second} inter-sample spacing per writer pair (2-2-2-2) and varying per-sample deadlines. Error margins denote one standard deviation.}
\label{fig:results:aggregation}
\end{figure*}

\section{Results}
\label{sec:results}
Since StreamRTPS is implemented as an extension of EmbeddedRTPS, we report all bandwidth and latency reductions relative to unmodified EmbeddedRTPS as the primary baseline. FastDDS numbers are included for reference.
We note that EmbeddedRTPS exhibits higher absolute transmission latencies than FastDDS across all payload sizes, likely due to architectural differences in EmbeddedRTPS's internal thread pool on the sender and receiver side.
StreamRTPS does not increase latency relative to EmbeddedRTPS, confirming that the overhead stems from the baseline implementation, not from our extensions.
The relative bandwidth savings reported in this section are a function of the payload-to-header ratio on the wire, which is governed by the RTPS message and submessage layout and is therefore independent of a specific RTPS implementation. Because FastDDS, EmbeddedRTPS, and other compliant implementations all emit the same \SI{44}{\byte} of RTPS header per packet, we expect the header-driven savings to transfer to other implementations.

\begin{figure*}[!t]
\centering
\makebox[\textwidth]{%
\begin{minipage}[t]{0.48\textwidth}
\vspace{0pt}
\centering
Heartbeat Suppression Bandwidth\par\medskip
\includegraphics[width=\textwidth]{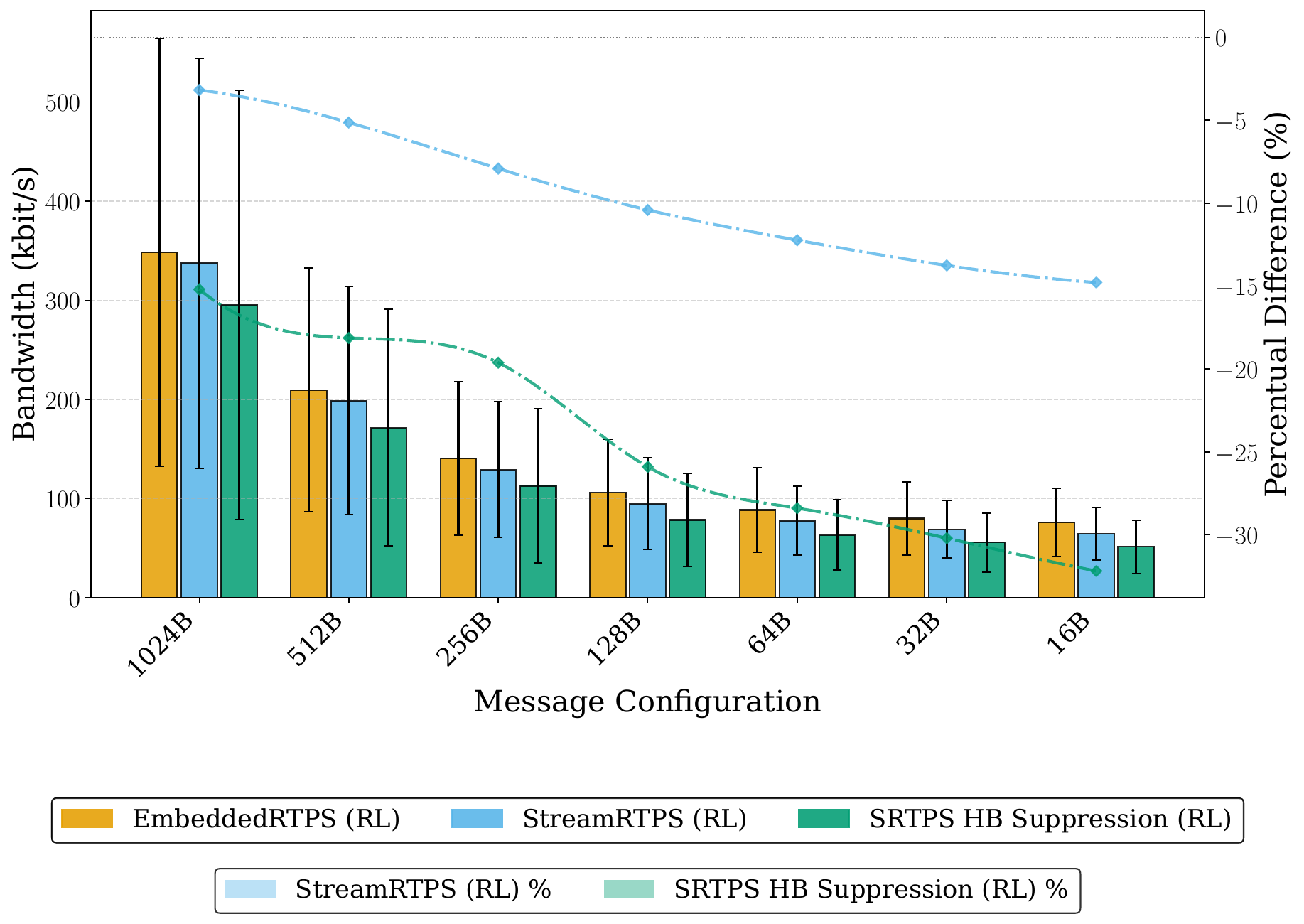}
\caption{Bandwidth with heartbeat suppression for varying payload sizes in a 4:2-4:2 topology with reliable transport. Error bars denote one standard deviation.}
\label{fig:results:heartbeat}
\end{minipage}%
\hfill%
\begin{minipage}[t]{0.49\textwidth}
\vspace{0pt}
\centering
Time-to-Retransmission under \SI{1}{\percent} Loss\par\medskip
\includegraphics[width=\textwidth]{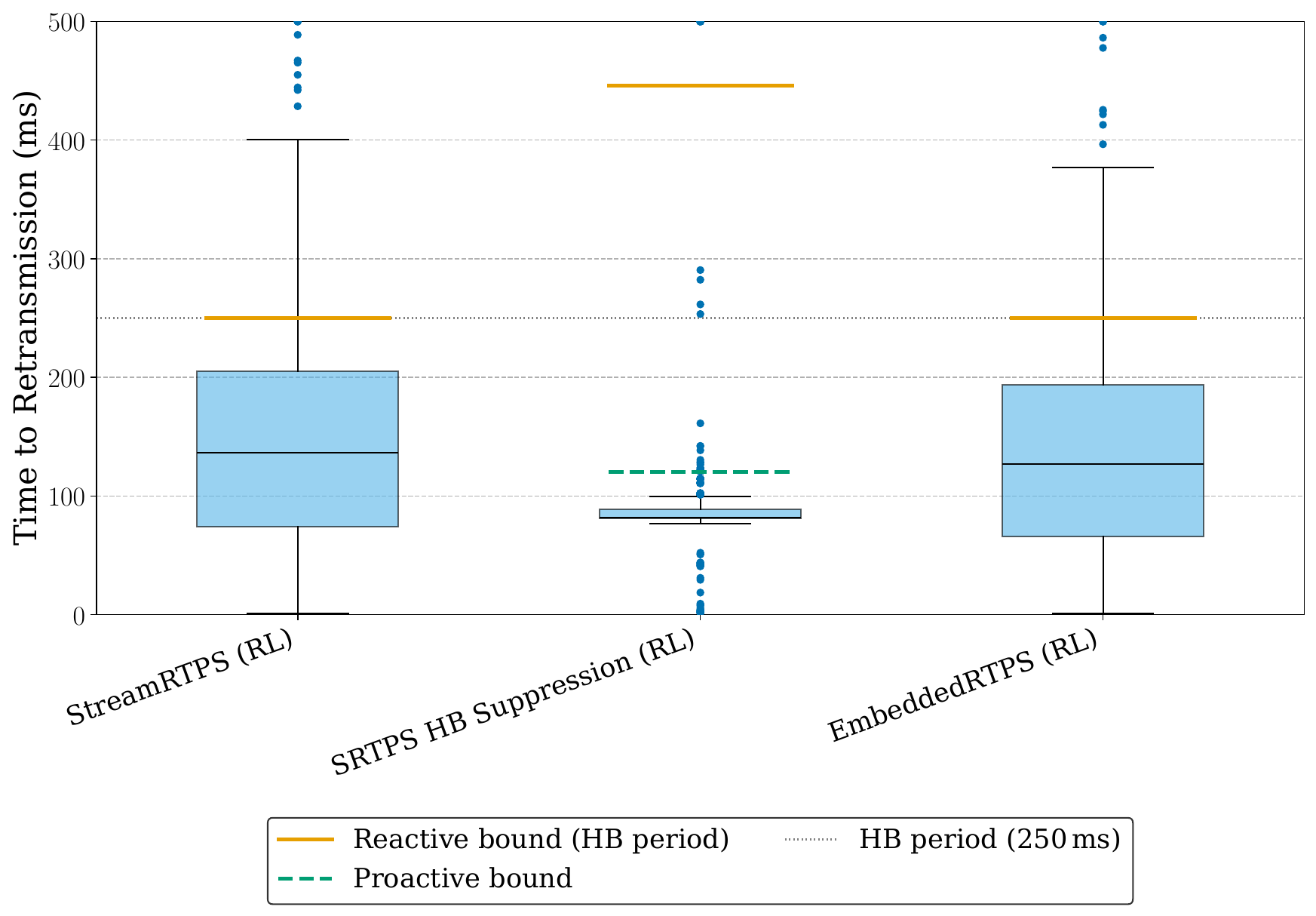}
\caption{Time-to-retransmission distribution under \SI{1}{\percent} induced packet loss for \SI{256}{\byte} payloads.}
\label{fig:results:retransmission}
\end{minipage}%
}
\end{figure*}

\begin{table}[b]
\centering
\caption{StreamRTPS bandwidth vs.\ send frequency for \SI{128}{\byte} payloads in a 4:2-4:2 topology.}
\label{tab:freq-sweep}
\begin{tabular}{lcc}
    \toprule
    \textbf{Frequency} & \textbf{EmbeddedRTPS} & \textbf{SRTPS} \\
    \midrule
    \multicolumn{3}{l}{\textit{Mean Bandwidth $B$ [\SI{}{\kilo\bit\per\second}]}} \\
    \SI{10}{\hertz}  & 75.3   & 66.7 (\textbf{\SI{-11.4}{\percent}}) \\
    \SI{25}{\hertz}  & 96.7   & 82.6 (\textbf{\SI{-14.5}{\percent}}) \\
    \SI{50}{\hertz}  & 131.3  & 108.7 (\textbf{\SI{-17.3}{\percent}}) \\
    \SI{75}{\hertz}  & 164.5  & 133.6 (\textbf{\SI{-18.8}{\percent}}) \\
    \SI{100}{\hertz} & 193.9  & 156.1 (\textbf{\SI{-19.5}{\percent}}) \\
    \bottomrule
\end{tabular}
\end{table}

\subsection{Stream-based transport}
\label{sec:results:streams}
Using our stream transport, we achieved a maximum bandwidth reduction of \SI{27.9}{\percent} compared to EmbeddedRTPS for \SI{16}{\byte} messages.
As noted in \cref{sec:intro}, small payloads amplify the relative cost of header overhead, which is apparent in this result.
The reduction decreases with payload size, reaching \SI{4.4}{\percent} for \SI{1024}{\byte} messages.
\Cref{fig:results:streams} presents the measured bandwidths and transmission latencies.
In the evaluated topologies, discovery uses multicast while user data is delivered via unicast between matched endpoints. Consequently, \SI{100}{\percent} of data-path bytes, which dominate steady-state traffic, use stream headers. Deployments relying on multicast data distribution (e.g., 1-to-many broadcast topics) would retain standard RTPS for those flows.

Compared to EmbeddedRTPS, the transmission latency is not adversely affected by the stream mechanism.
Although minor variations of a few percent are observed, these fall within measurement uncertainty.
Stream negotiation adds one additional round-trip to the discovery phase.
In our measurements, this overhead remains within the variance of EmbeddedRTPS's standard SPDP/SEDP discovery procedure and could not be isolated as a distinct latency contribution.

To examine the effect of send rate, we additionally swept the transmission frequency from \SIrange{10}{100}{\hertz} for \SI{128}{\byte} payloads, as shown in \cref{tab:freq-sweep}.
The bandwidth reduction increases with frequency, from \SI{11.4}{\percent} at \SI{10}{\hertz} to \SI{19.5}{\percent} at \SI{100}{\hertz}.
This is expected as at higher send rates, more packets per second are transmitted, each benefiting from the reduced stream header, so the absolute header savings scale linearly with frequency.

These results are measured under best-effort transport. For reliable transport, the relative savings are lower: the Reliable stream header (\SI{12}{\byte} vs.\ \SI{2}{\byte} Simple) reduces per-packet savings, and control traffic such as heartbeats and \acp{NACK} is unaffected by stream headers, limiting the overall reduction.
This motivates our heartbeat suppression mechanism (\cref{sec:results:hb_suppression}), which targets this remaining control overhead.
Overall, StreamRTPS achieves increased bandwidth efficiency at no latency cost, with savings that scale with send frequency.
CPU and memory usage remained within measurement noise of EmbeddedRTPS across all payload sizes (mean CPU delta \SI{-0.55}{\percent}, mean RAM delta \SI{-0.53}{\percent}), confirming no measurable host-side resource overhead.

\subsection{Cross-Writer Aggregation}
\label{sec:results:aggregation}
Aggregating messages yielded another increase in bandwidth efficiency at predictable latency cost. 
We evaluated our aggregation approach in an 8:1-8:1 topology to increase the count of co-located publishers. 
This topology allows us to examine scaling behavior of our aggregation approach when aggregating increasing message counts.
We configured a transmission timing of 2-2-2-2, resulting in 2 publishers per machine sending samples each \SI{10}{\milli\second} within the \SI{40}{\milli\second} transmission window given by the \SI{25}{\hertz} transmission frequency.
\cref{fig:results:aggregation} illustrates measured results in relation to the configured per-sample deadline.
Aggregation for a \SI{2}{\milli\second} deadline resulted in a bandwidth reduction of \SI{10.2}{\percent} for \SI{128}{\byte} samples compared to StreamRTPS and \SI{30}{\percent} compared to EmbeddedRTPS.

For a \SI{2}{\milli\second} deadline, the transmission latency increased to \SI{2.95}{\milli\second} and allowed the aggregation of the first two samples. 
The significant latency increase is expected, as  our aggregation approach holds the sample with the earliest deadline until deadline expiration.
Consequently, a \SI{2}{\milli\second} deadline per-sample leads to at least one sample with more than two milliseconds transmission delay, also shown in \cref{fig:results:aggregation} through the linearly increasing maximum latencies in relation to their deadline. 
The mean varies non-monotonically because the remaining samples within the aggregation window arrive at intermediate offsets and shift the average.
In future work we plan to consider other approaches, which use system knowledge to remove the need to hold samples to deadline.

Additional bandwidth efficiency gains occurred when more samples could be aggregated, specifically when crossing the \SI{10}{\milli\second} and \SI{20}{\milli\second} thresholds.
At each \SI{10}{\milli\second} increment another two messages can be added to the aggregation window.
These thresholds reduced bandwidth usage by an additional \SI{4.6}{\percent} and \SI{1.5}{\percent} respectively, indicating diminishing returns for higher sample aggregations.

In conclusion, aggregation of a small number of samples with similar release times and short deadlines appears to be effective. 
However, aggregating beyond two samples in our case with deadlines exceeding \SI{10}{\milli\second} is not realistic and yields diminishing bandwidth returns at disproportionate latency costs.
RAM usage remained unchanged across all configurations, while aggregation reduced mean CPU usage by \SI{8.5}{\percent} relative to EmbeddedRTPS due to reduced per-packet networking overhead.

\subsection{Heartbeat Suppression}
\label{sec:results:hb_suppression}
We evaluated heartbeat suppression in a 4:2-4:2 topology using reliable transport using the larger StreamHeader, including sequence numbers.
\Cref{fig:results:heartbeat} shows bandwidth across payload sizes: suppression yielded a \SI{22.7}{\percent} reduction for \SI{16}{\byte} messages and \SI{19.7}{\percent} for \SI{128}{\byte} compared to StreamRTPS.
As this mechanism only modifies the reliability control path, transmission latency was not affected, and CPU and RAM usage remained indistinguishable from both EmbeddedRTPS and StreamRTPS across all payload sizes.

To validate loss recovery, we introduced \SI{1}{\percent} i.i.d.\ packet loss at the UDP transport level for \SI{256}{\byte} payloads.
All affected samples were recovered through retransmission and none failed to be delivered to the application.
\Cref{fig:results:retransmission} shows the time to retransmission, which we define as the interval from original send to retransmission after loss detection.
With heartbeat suppression, the mean time to retransmission was \SI{124.1}{\milli\second}, compared to \SI{148.6}{\milli\second} for EmbeddedRTPS and \SI{149.8}{\milli\second} for StreamRTPS without suppression.
Rather than degrading loss recovery, suppression improves it: the reader's proactive loss detection (\cref{sec:method:traffic_implied}) fires at \SI{120}{\milli\second}, well before the standard heartbeat period of \SI{250}{\milli\second}.

The parameters in \cref{tab:hb-params} were selected to balance suppression against safety margin: $\sigma^2_{\max} = \SI{400}{\micro\second\squared}$ (corresponding to a standard deviation of \SI{20}{\micro\second}) rejects sample streams with excessive jitter.
Heartbeat suppression thus offers significant bandwidth savings while simultaneously improving loss recovery latency for periodic transmissions, making it well-suited for DDS deployments using reliable transport with periodic transmission.

\section{Conclusion}
\label{sec:conclusion}
This paper presented three backward-compatible extensions to the RTPS wire protocol: stream-based header replacement, deadline-aware message aggregation, and predictive heartbeat suppression.
Experimental evaluation demonstrates that stream headers alone reduce bandwidth consumption by up to \SI{27.9}{\percent} for small payloads and \SI{11.8}{\percent} for \SI{256}{\byte} messages compared to EmbeddedRTPS, without introducing measurable latency penalties.
As shown in \cref{fig:motivation_result}, our extensions achieve competitive bandwidth efficiency with compact wire formats while preserving RTPS interoperability.
StreamRTPS is particularly relevant to any DDS deployment with small or high-rate messages, especially periodic control traffic and small sensor samples.
Heartbeat suppression also yields reductions for reliable transport of up to \SI{22.7}{\percent}, and is effective for the periodic traffic patterns prevalent in industrial and control workloads.
Message aggregation offers additional savings at the cost of deadline-dependent increased transmission latency and should be used selectively in suitable network environments.
Current limitations include the restriction to unicast for stream transport and the requirement for QoS deadlines to drive aggregation, which precludes its use for traffic without configured deadlines. Future work will address these limitations.

\printbibliography

\end{document}